\documentstyle[epsfig,graphicx,12pt]{article}
\begin{document}

\title{ {\bf Estimate of Crossed-Boson-Exchange Contributions \\ to the Binding 
Energy of Two-Body Systems }}
\author{ Gwena\"{e}l Le Gorrec 
and Bertrand Desplanques$^a$\thanks{{\it E-mail 
address:}  
desplanq@isn.in2p3.fr}\\
$^a$Institut des Sciences Nucl\'eaires (UMR CNRS/IN2P3--UJF),  
\\ F-38026 
Grenoble Cedex, France
}

\maketitle

\begin{abstract}
Binding energies calculated from using the Bethe-Salpeter 
equation in the simplest ladder approximation significantly differ 
from those obtained in the non-relativistic standard instantaneous 
approximation. While they should {\it a priori} be better, they turn 
out to be further away from an exact calculation in the case 
of scalar neutral particles or from experiment in the case 
of the Coulomb interaction. Part of the discrepancy is due 
to the omission in the interaction kernel of contributions 
corresponding to crossed-boson-exchange diagrams. The role 
of these contributions is examined numerically, using a simple 
approximation. The sensitivity to both the coupling constant 
and the mass of the exchanged boson is considered. 
\end{abstract} 

\newpage
\section{Introduction}
In calculating binding energies of a two-body system, it is a current 
practice to rely on a meson-exchange-mediated interaction in the 
instantaneous approximation. Apparent support for this approach 
in the non-relativistic domain comes from the success of the Coulomb 
interaction in explaining atomic spectra. In the case of a massive 
neutral scalar boson exchange, the standard Yukawa potential provides 
binding energies that are half-way from those given by a 
field-theory-motivated calculation \cite{NIEU}.  Curiously, in both 
cases, calculations based on the Bethe-Salpeter equation \cite{BETH} 
in the  ladder approximation, which should represent an improvement 
over these non-relativistic calculations, are not doing so well. 
The binding energies so obtained, for a zero-mass boson case 
(Wick-Cutkosky model \cite{WICK,CUTK}) or a massive one, 
are much lower (see Refs. \cite{SILV,BILA} and \cite{NIEU,PHIL} respectively). 

The discrepancy has been examined in different papers \cite{PHIL,AMGH,AMGH2} 
where it was shown that the standard instantaneous approximation, 
which leads to the Coulomb or Yukawa potentials, is too a drastic 
approximation. It neglects retardation effects in relation with 
the fact that, part of the time, the two constituents are accompanied 
by the bosons they are exchanging. Simplifying a little, one can 
say that, during this time, the constituents do not interact with 
each other as far as one relies on the ladder approximation. 
The force between the constituents is thus effectively  reduced, 
hence smaller binding energies. However, if one imagines that 
these constituents are allowed to interact and that this interaction 
is the same as in absence of in-flight bosons, one can expect 
the effective interaction between the constituents to be transparent 
to these in-flight bosons, recovering in this case the standard 
instantaneous-approximation-potential \cite{TODO,BREZ,FRIA,GROS,NEGH,ITZY}. 
This supposes to incorporate in the interaction  the contribution of 
crossed-boson exchanges. It also supposes that the exchanged bosons 
carry neither spin nor charge, otherwise the interaction between 
the constituents would differ, depending whether there are 
in-flight bosons or not. 

There are rather few calculations incorporating the contribution 
of crossed-boson exchange diagrams. Moreover, they essentially 
involve two-boson exchange. This was considered in atomic physics 
\cite{LEPA} or for the nucleon-nucleon interaction \cite{PARI,BONN}. 
Within theoretical models, this was also considered for scalar 
particles where their effect was found to almost remove the above 
retardation effect on the strength of the interaction \cite{THEU}. 
Later studies however revealed that  the meaning of these results 
was not quite clear due to bad convergence properties with respect 
to the order of diagrams under consideration. Part of the problem 
arises from the fact that, for some diagram, it is possible 
to incorporate more or less of the higher order contributions 
(with respect to the coupling constant). Thus, an alternative 
approach \cite{AMGH2} led to a significantly lower contribution 
from the two-crossed-boson exchange to the binding energy, 
leaving entire the problem of recovering at least the 
benchmark-binding-energies obtained in the instantaneous 
approximation. A similar conclusion could be drawn from results 
obtained by other approaches \cite{TJON}. 

In the present work, we will consider the contribution of higher-order 
crossed-boson-exchange diagrams. As exact calculations are practically 
unfeasible\footnote{unless to use methods of ref. \cite{NIEU}, but in that case 
no detailed information is available on the role of different orders for 
instance}, we will use some approximation which, actually, 
underlies for a part the  demonstrations tending to show that the 
crossed-boson-exchange allows one to recover the instantaneous 
potential in the case of spin-less and charge-less bosons  
\cite{TODO,BREZ,FRIA,GROS,NEGH,ITZY}. Taking for granted the 
above theoretical works, the main motivation of the present paper 
will be to illustrate numerically how the convergence to the 
expected result is achieved, depending on the coupling or 
on the mass of the exchanged boson. On the other hand, the above 
theoretical works are not so well known and we believe that it is 
not unnecessary to remember the role of the crossed-boson exchange 
in recovering the standard instantaneous-approximation-potential. 
This study is made at the lowest order in the inverse of the 
constituent mass.

The plan of the present paper is as follows. In the second part, we remind 
the expressions that we are using and their relation to a diagram 
expansion. The third part is devoted to the presentation 
of the results for the binding energies in terms of the coupling 
constant and the boson mass. This is done as a function of the order 
of the crossed diagrams that are taken into account. A short conclusion is given 
in the fourth section.

\section{Expressions used for the contribution of crossed diagrams to the 
interaction}
We first consider in this section the expression of the effective 
potential introduced to account for some retardation effects, 
going therefore beyond the standard instantaneous approximation 
\cite{AMGH,AMGH2}. In the 
following part, we start from this effective interaction to introduce 
a contribution that could globally account for crossed diagrams. 
How this is realized in practice is detailed.

\subsection{Effective interaction in the ladder approximation}  
\begin{figure}[htb]
\begin{center}
\mbox{ \epsfig{ file=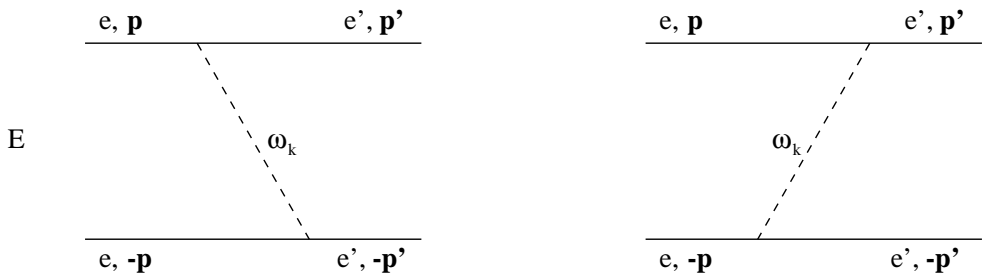, width=13cm}}  
\end{center}
\caption{Time-ordered single-boson-exchange contributions to 
the two-body interaction with indication of the kinematics 
in the center of mass.\label{time-ordered}}
\end{figure}

Our starting point for our study  is a single-boson-exchange 
contribution, field-theory motivated interaction, represented 
by the time-ordered diagrams represented in Fig. \ref{time-ordered}. 
 
Its expression is obtained from second-order perturbation theory.
In the center of mass, the two terms representing the contributions 
of the diagrams displayed in Fig. 1 are equal, providing 
some simplification and allowing one to write:
\begin{eqnarray}
V_E(\vec{p},\vec{p}\,')= 
\frac{g^{2}  \,}{\omega_{k}} \,  
\left( \frac{1}{E-\omega_{k}-e-e'}  \right) ,
\hspace{1cm} \nonumber \\
{\rm with} \;\;\;\;\;\;\; \omega_{k} = \sqrt{\mu^{2}+\vec{k}^{\,2}},\;\;\;\;
e= \frac{\vec{p}^{\,2}}{2\,m}, \;\;\;\;
e'=   \frac{\vec{p}\,'^{\,2}}{2\,m} ,
\label{potE}
\end{eqnarray}
$\vec{k},\, \vec{p},\, \vec{p}\,'$ representing respectively 
the momentum of the exchanged boson and of the constituents 
in the initial and final states. Anticipating on the non-relativistic 
character of the present work, the different energies appearing 
in the above interaction, Eq. (\ref{potE}), have been replaced by their 
non-relativistic 
counter-part. The energy dependence appearing at the denominator 
of the meson propagator, which stems from the field-theory 
character of the approach, is nevertheless fully kept. 
As shown in \cite{PHIL,AMGH,AMGH2}, it is the key ingredient that 
allows one to approximate results from the Bethe-Salpeter equation 
in the laddder approximation while providing departure to the 
instantaneous-approximation results. These last ones are recovered 
when the term at the denominator, $E-e-e'$, is neglected.

Consistently with neglecting relativistic corrections, 
the above interaction is used with the following 
non-relativistic equation:
\begin{equation}
\Big(\frac{\vec{p}\,^2}{m} - E\Big) \, \psi(\vec{p}\,) =
-\int \frac{d\vec{p}\,'}{(2\pi)^3} \; 
\Big(V_E(\vec{p},\vec{p}\,')+ \dots\Big) \; \psi(\vec{p}\,'),
\label{schE}
\end{equation}
where dots stand for higher order contributions. 
The use of an energy-dependent-interaction raises some problem 
such as non-orthogonality of the solutions. This energy dependence 
can be removed by a transformation quite similar in its spirit 
to the Foldy-Wouthuysen one, at the price of dressing the original 
degrees of freedom to get effective ones (see also refs. \cite{FST,OKUB}). 
It also results an 
effective interaction. At the lowest order $(\frac{1}{m})^0$, 
this interaction can be obtained by replacing the energy-dependent 
term at the denominator in Eq. (\ref{potE}), $E-e-e'$, 
by the interaction itself, obtaining in this way a self-consistency 
equation for the effective potential. In configuration space, 
this equation reads:
\begin{equation}
V_{eff,sc}(r)=-g^{2} \; \frac{1}{2\pi^{2}}
  \int \frac{dk\; k^{2} j_{0}(kr)}{\omega_{k}
  \Big(\omega_{k} -V_{eff,sc}(r)\Big)}.
  \label{sc}
\end{equation}
This potential may also be used in a non-relativistic equation: 
 \begin{equation}
\left( V_{eff,sc}(r)+\frac{\vec{p}^{\,2}}{m} -E \right) \phi(r)=0,
\label{sch}
\end{equation}
which is a typical Schr\"{o}dinger one.  
Its solutions, which are now orthogonal, will provide the binding 
energies of interest here. It has been shown that it retains most 
of the genuine character related to the energy dependence of 
the original interaction, Eq. (\ref{potE}) \cite{AMGH,AMGH2}. 
The change from the function, $\psi(\vec{p}\,)$, 
in Eq. (\ref{schE}) to the function, $\phi(r)$, in 
Eq. (\ref{sch}) mainly reminds that they correspond to different 
formalisms (energy-dependent and energy-independent), the change 
from momentum to configuration space being unimportant here.
In the limit where 
one could neglect the interaction term in the denominator at 
the r.h.s. of Eq. (\ref{sc}), the effective interaction would 
identify to the standard instantaneous interaction:
\begin{equation}
V_{0}(r)=-g^{2} \,    \frac{ e^{-\mu r} }{4\pi r}.
\label{nr}
\end{equation}

\subsection{Beyond the ladder approximation}

In deriving the interaction, Eq. (\ref{potE}), a standard approach 
has been used. This one does not incorporate possible interaction 
between the two constituents while a boson is exchanged. To get 
some guess on its effect, a first approach would consist in  adding 
this interaction to the denominator of the meson propagator in this 
equation, as part of the energy in the intermediate state, 
or, better, in Eq. (\ref{sc}). From this one, the following expression 
is  immediately obtained:
\begin{equation}
V_{eff,sc}(r)=-g^{2} \;  \frac{1}{2\pi^{2}}
  \int \frac{dk \; k^{2} j_{0}(kr)}{\omega_{k}
  \Big(\omega_{k} -V_{eff,sc}(r)+V_{eff,sc}(r)\Big)}=V_0(r).
  \label{scsc}
\end{equation}
The above result supposes that the interaction with and without 
in-flight bosons are the same, which implies that the exchanged 
bosons couple to the constituents in a scalar way (no spin, no charge). 
The effect of the extra interaction between the constituents 
is depicted on the left diagram in Fig. 2. Replacing the bubble representing 
this interaction by a single-boson-exchange immediately suggests 
that this effect actually involves crossed-boson-exchange diagrams as can be 
seen on the r.h.s. of this figure.

\begin{figure}[htb]
\begin{center}
\mbox{ \epsfig{ file=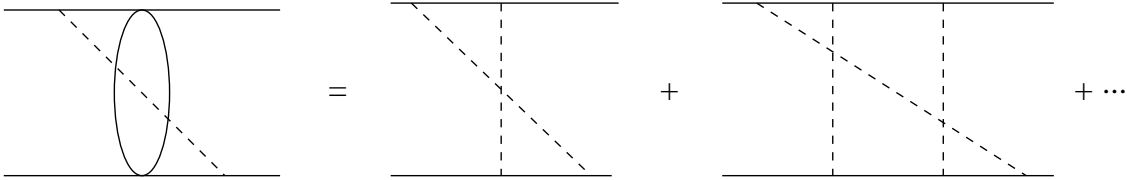, width=15cm}} 
\end{center}
\caption{Schematic representation of the interaction between 
the constituents while a boson is exchanged (left diagram) and when 
this one is represented by crossed diagrams with two and three bosons 
exchanged.}
\end{figure}

This feature can be checked by expanding the denominator in  
Eq. (\ref{scsc}) with respect to the effective interaction that 
has just been introduced:
\begin{eqnarray}
V_{eff,sc}(r)&=&-g^{2} \;  \frac{1}{2\pi^{2}}
  \int \frac{dk \; k^{2} j_{0}(kr)}{\omega_{k}\;\Big(\omega_{k} 
-V_{eff,sc}(r)\Big)}\;
  \nonumber \\
 & & \left( 1 + \frac{-V_{eff,sc}(r)}{\omega_{k} -V_{eff,sc}(r)}
  + \Big(\frac{-V_{eff,sc}(r)}{\omega_{k} -V_{eff,sc}(r)}\Big)^2
  +\dots \right).
  \label{exp}
\end{eqnarray}
It is easy to see that, apart from the first term representing 
a single-boson exchange, there is a one to one 
correspondance between this expansion and diagrams shown in Fig. 2. 
Symbolically, the expression of the nth order contribution reads:
\begin{equation}
\Delta\;V^{(n)}=\frac{-g^2}{E-H_0}\;\Big(V\;\frac{1}{E-H_0}\Big)^n,
\label{deltaV}
\end{equation}
where the quantity, $E-H_0$, can be written as $E-e-e'-\omega_k$. 
The quantity $V$ represents the (effective) interaction between 
the two constituents. Assuming that $E-e-e'=V_{eff,sc}(r)$, 
as we assumed for the derivation of the effective interaction 
in the ladder approximation \cite{AMGH2}, one gets:
\begin{equation}
\Delta\;V^{(n)}(r)=-g^{2} \;  \frac{1}{2\pi^{2}}
\int \frac{dk \; k^{2} j_{0}(kr)}{\omega_{k} \;\Big(\omega_{k} 
-V_{eff,sc}(r)\Big)} 
\; \Big(\frac{-V_{eff,sc}(r)}{\omega_{k} -V_{eff,sc}(r)}\Big)^n.
\label{deltaVr}
\end{equation}
This expression is identical to the nth order term in the expansion 
given in Eq. (\ref{exp}). 
We now comment on what the above expansion accounts for. At first sight, 
it seems to only involve part of the crossed-boson-exchange diagrams. 

\begin{figure}[htb]
\begin{center}
\mbox{ \epsfig{ file=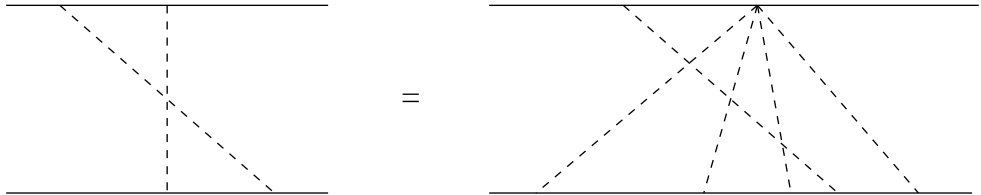,  width=13cm}} 
\end{center}
\caption{Actual time-ordered contributions (r.h.s.) approximately accounted for 
by the first-order crossed diagram (l.h.s.). Notice that the very right diagram, 
which has a non-crossed character, is required to factorize the expression of an 
instantaneous interaction when added to the other ones.}
\end{figure}

First of all, it is noticed that expansion whose general term 
is given by Eq. (\ref{deltaV}) contains the most singular terms.
In  the zero-mass boson case and in absence of the effective 
interaction at the denominator, the nth order contribution 
contains a factor $\omega_k^{-n}$ and is therefore 
infra-red diverging. This divergence is hopefully regularized 
by the effective interaction. Secondly, the interaction $V$ at 
the numerator accounts for many more diagrams than inferred 
from Fig. 2, including some non-crossed diagrams that 
have a small contribution. This 
can be checked in the small coupling limit and has been done 
explicitly in ref. \cite{AMGH} for two- and three-boson exchange. 
Diagrams that are actually included in the two-boson-exchange case 
are shown in Fig. 3. The contribution of each diagram in this 
figure is rather complicated but great simplication occurs when 
they are summed up, with the result that propagators relative 
to each meson factorize. Lastly, we stress that the developments 
presented here suppose that the effective interaction can be 
approximated by a local one. This makes it possible to derive 
relatively simple expressions and facilitates calculations. 
Going beyond would require tremendous work. Moreover, as the 
corrections are of order $(\frac{1}{m})$ in the minimal case, 
it is likely that one should also consider for consistency 
relativistic corrections of same order. This involves $\sqrt{m/e}$ 
factors in the interaction, Eq. (\ref{potE}), and many boson-exchange 
contributions involving Z-type diagrams.

\section{Results}

\begin{figure}[htb]
\begin{center}
\mbox{ \epsfig{ file=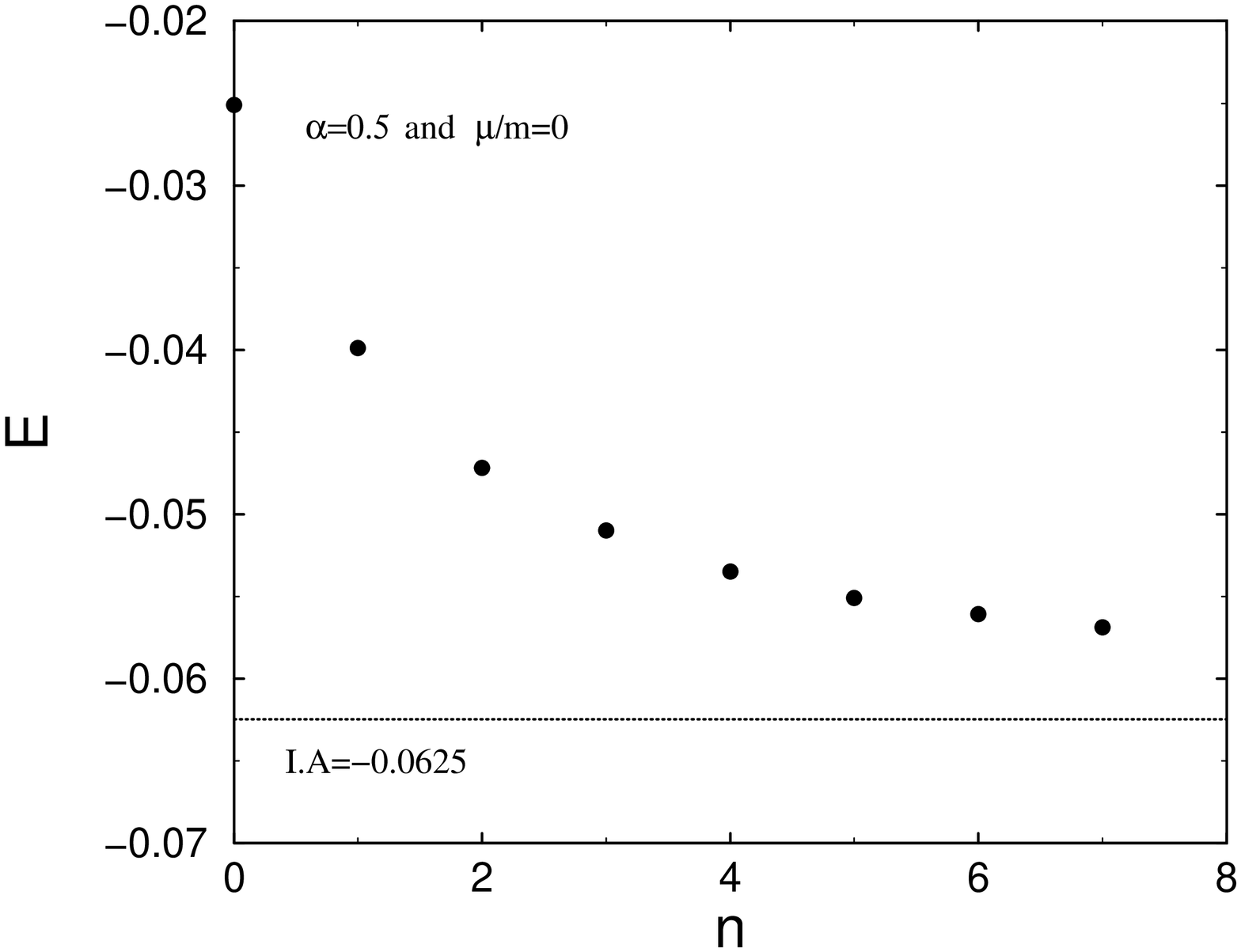,angle=270, width=6.6cm}
\hspace{1cm} \epsfig{ file=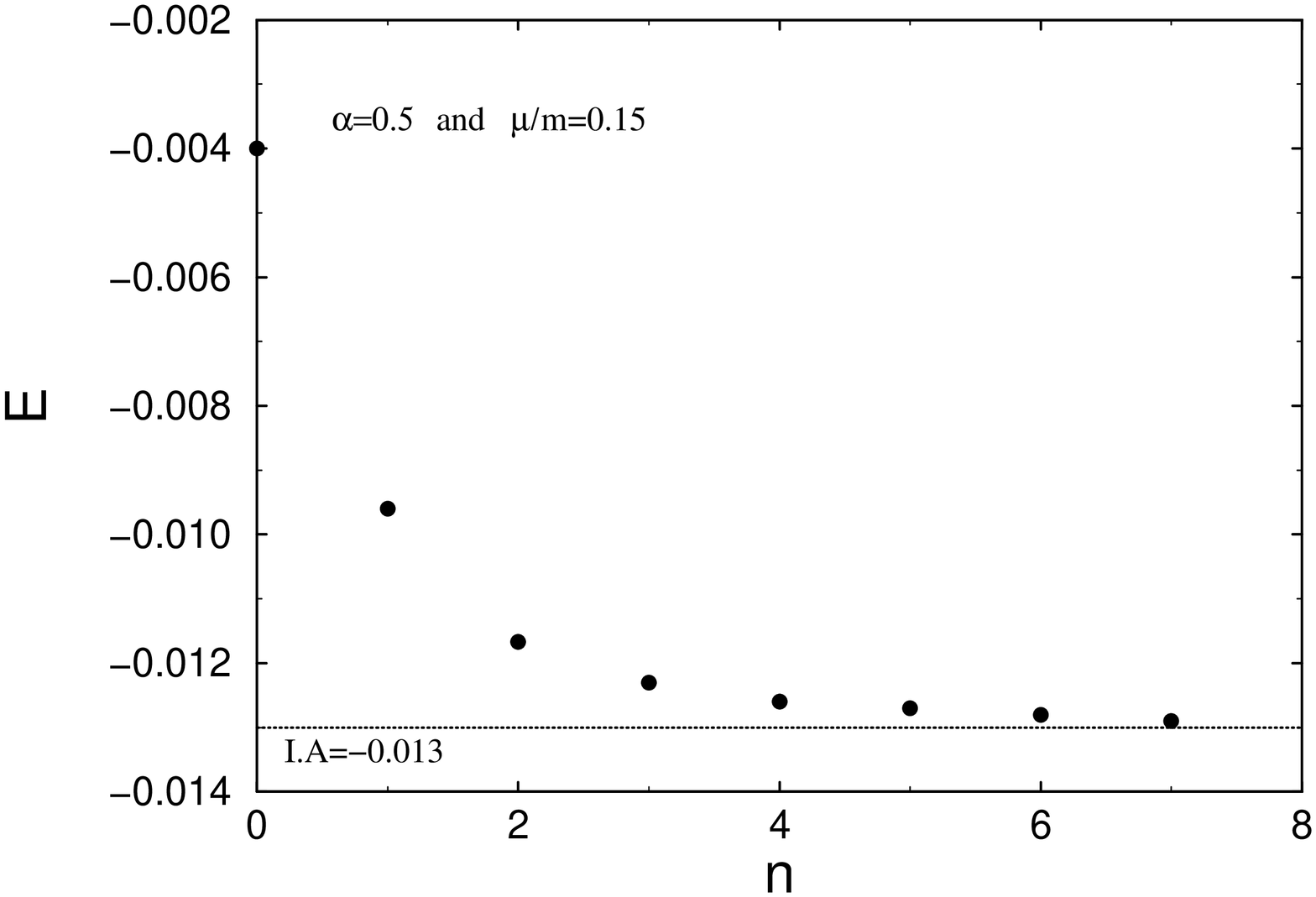,angle=270, width=7.4cm}} 
\end{center}
\caption{Contribution of crossed diagrams to the binding energy 
as a function of the number of exchanged bosons, $n+1$: results 
for the coupling constant, $\alpha=0.5$ and two boson masses, $\mu=0$  
and  $\mu=0.15\,m$. The asymptotic result (denoted I.A.) is represented 
by the horizontal dashed line. The energy unit is the constituent mass, $m$.}
\end{figure}

\begin{figure}[htb]
\begin{center}
\mbox{ \epsfig{ file=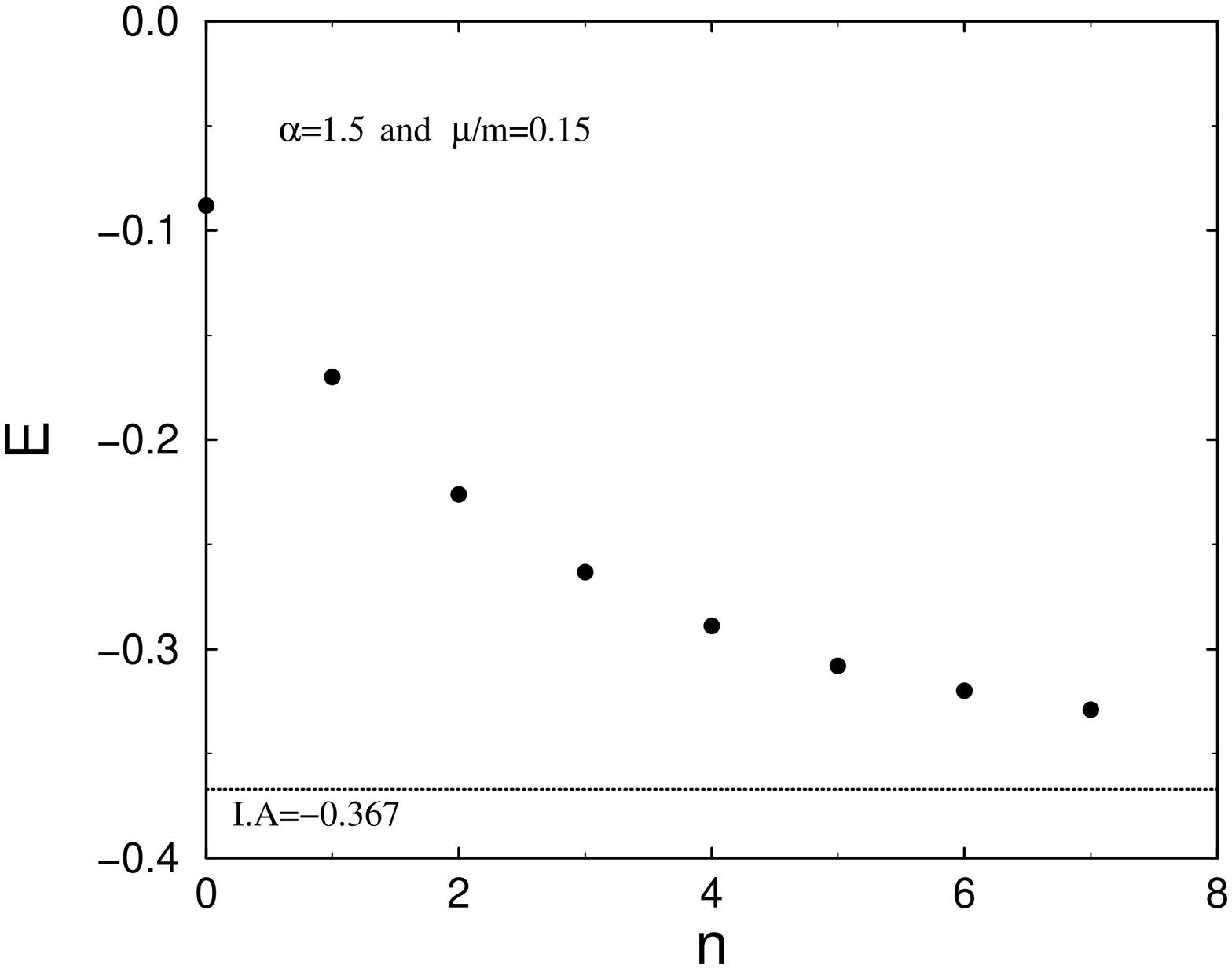, angle=270, width=6.9cm} 
\hspace{1cm} \epsfig{ file=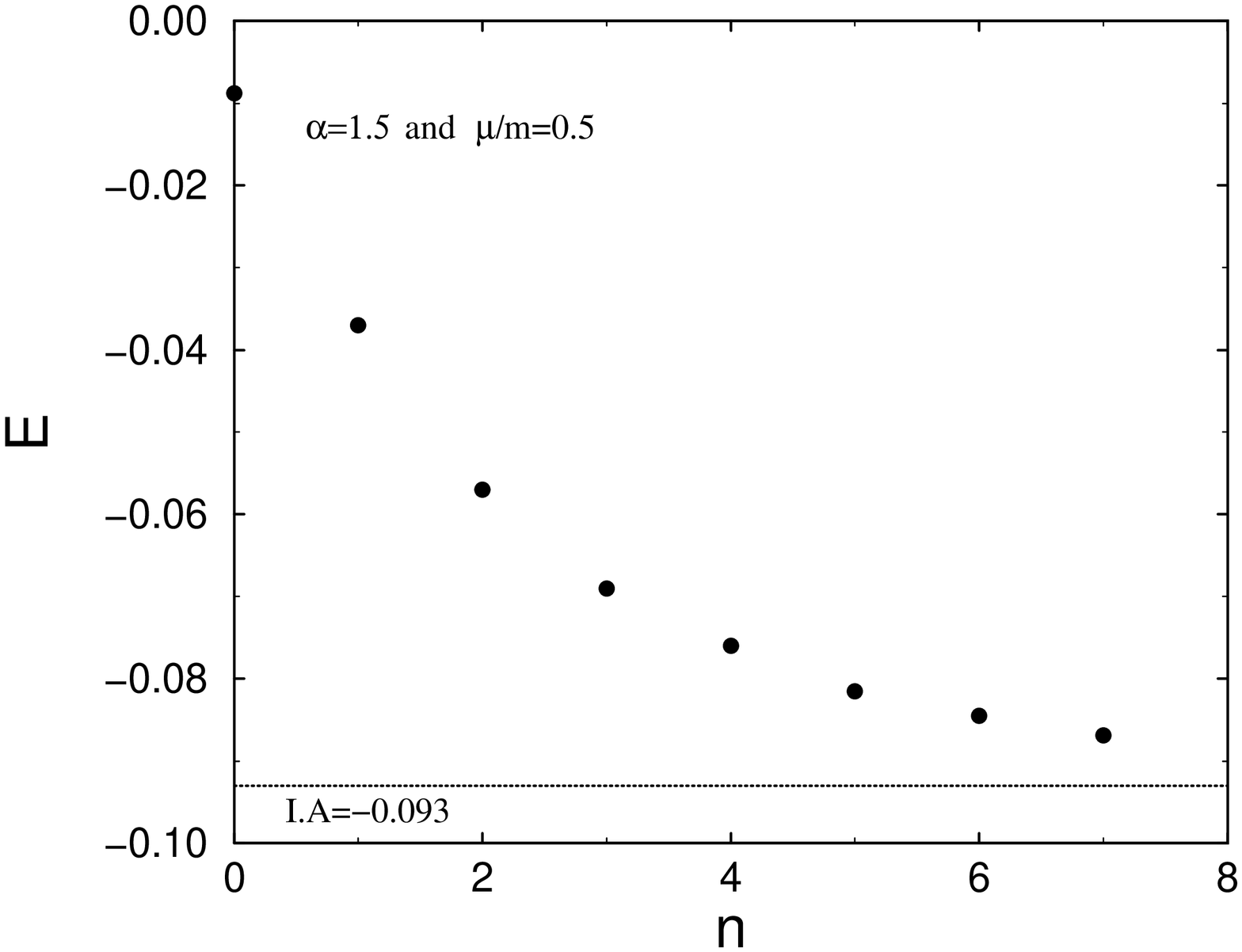, angle=270, width=7.1cm} }
\end{center}
\caption{Same as for Fig. 4  for the coupling constant, $\alpha=1.5$,  and two  
boson masses, $\mu=0.15\,m$ and  $\mu=0.5\,m$.}
\end{figure}

In calculating the contribution of each crossed diagram, a first 
possibility is to calculate the contribution of each term in the 
expansion of the effective potential, Eq. (\ref{exp}), assuming 
that the problem is solved and that, according to Eq. (\ref{scsc}), 
this self-consistent effective potential, $V_{eff,sc}(r)$, is equal 
to the instantaneous one, $V_0(r)$. This procedure however provides 
results whose interpretation is somewhat ambiguous. While the 
contribution of a given crossed diagram is obtained, the contribution 
due to the kinetic energy contains contributions at all orders 
when calculated with wave functions that are solutions of the problem. 
As a result, the sum of the kinetic energy and the first order term, 
for instance, may not evidence any binding. We will therefore proceed 
differently.

To get a better insight on the role of the different crossed diagrams, 
we will truncate the expansion of the effective potential, Eq. (\ref{scsc}), 
at some order, determine the associated effective potential, insert this 
potential in the Schr\"{o}dinger equation, Eq. (\ref{sch}), and look for  
the  corresponding binding energy. In this way, the kinetic energy 
is calculated consistently with the employed potential. We thus obtain 
a series of binding energies which ultimately should converge to the 
binding energy obtained with the standard instantaneous potential. 
Our interest is in the convergence of this series to the expected value, 
depending on the strength of the coupling constant, or on the mass of the 
exchanged boson.

For the coupling constant, denoted $\alpha=\frac{g^2}{4\,\pi}$ 
in analogy with the QED one, we consider two values, $\alpha=0.5$ 
and $\alpha=1.5$.  In each case, two boson masses are considered, 
so that the sensitivity to this mass can be studied. In the first case, 
we included the boson masses, $\mu=0$ and $\mu=0.15\,m$, 
the first of them being  appropriate to the study of a 
Coulomb-like interaction. In the other case, we included 
the masses,  $0.15\,m$ and $0.5\,m$,  that have been used in various works 
dealing with the scalar-particle model. The first of them is identical 
to one of the values  introduced for the coupling, $\alpha=0.5$, 
so that to also provide insight on the sensitivity of results 
to this coupling. The results are presented in two figures, Fig. 4 
for $\alpha=0.5$ and Fig. 5 for $\alpha=1.5$.

From examination of the two parts of Fig. 4, it is seen that the convergence 
to the asymptotic value is slower when the boson mass decreases. The same 
observation holds for results presented in Fig. 5. This feature, which is 
perhaps counter-intuitive, can be explained by the fact that, with a lower 
boson mass, the force extends to larger distances, making the effect bigger. 
There is an evident relationship with the increasing singular character 
of the terms of the expansion of the effective interaction, Eq. (\ref{exp}), 
when the effective interaction at the meson propagator is neglected 
and the boson mass goes to zero. 

The comparison of results of Figs. 4 and 5 for the same boson mass, 
$\mu=0.15\,m$, shows that the convergence is slower when the coupling 
strength increases. This result is much less surprising as the role 
of higher order diagrams in the interaction is expected to increase 
with the strength of the coupling. Thus, altogether, it turns out 
that the more or less rapid character of the convergence to the 
asymptotic value of the binding energy is governed by the global 
interaction. This one combines both the intensity of the coupling 
and the range of the force, the interaction being bigger when it 
extends to larger distances. Roughly, the slowness of the convergence 
increases with the binding energy. 

While the first-order crossed diagram provides more than 2/3 
of the missing binding energy in the  less bound case ($E=-0.013\,m$, 
right part of Fig. 4), in the other cases, it hardly provides 
a third of it  
and one has to include the exchange of three crossed bosons 
to reach a half. In these cases, the interaction of crossed diagrams 
up to order 7 does not allow one to approach the asymptotic value 
at better than 7-8\%. 

The results obtained here are in complete agreement with those obtained 
in refs.  \cite{AMGH2,TJON}. They do not support however those, 
much larger, of ref. \cite{THEU}, based on a dispersion calculation 
of the crossed-box diagram. This discrepancy is interesting. It mainly 
reveals that individual contributions can be quite large, but these ones 
may be dramatically reduced by higher order effects arising from 
the appearance of the effective interaction at the denominator of the boson 
propagator, $(\omega_{k} -V_{eff,sc}(r))^{-1}$. This has the advantage 
to stabilize the corresponding contribution of each diagram in such 
a way that they  always have the same sign, making the expansion safer. 
In the other approach  \cite{THEU}, it is likely that higher order terms 
in the interaction will appear  with different signs, making difficult 
to make an accurate prediction. 

\section{Conclusion} 
It is not so well known that the validity of the standard instantaneous 
interaction largely relies on the cancellation of two effects and essentially 
supposes a scalar coupling of spin- and charge-less bosons to constituents. 
The first effect is a renormalization of the instantaneous interaction, 
which gets effectively reduced due to the absence of interaction between 
the two constituents while they are exchanging bosons. This effect is in 
agreement with that one found with the Bethe-Salpeter equation in the ladder 
approximation or an energy-dependent interaction, which, both, represent 
an improvement upon the standard instantaneous interaction. The second 
effect is mainly due to the contribution of crossed diagrams, which precisely 
account for the interaction between these constituents while bosons are 
exchanged. Our study was aimed to study the role of these higher order 
contributions in calculating binding energies.

Technically, we assumed that the above cancellation holds exactly. We could 
thus look at the convergence to the expected  binding energies, depending 
on the mass of the exchanged boson or on the size of the coupling constant. 
Except for very low binding energies (1\% of the constituent mass), 
we found that the convergence is rather slow. This feature is enhanced 
when the strength of the coupling increases or when the boson mass decreases. 
For usual coupling strengths employed in the domain of strongly interacting 
systems, retaining crossed diagrams up to order 6 does not allow one 
to approach the expected result with an accuracy much better than 10\%. 

The above results have been obtained by making some approximations, 
neglecting in particular non-locality effects and/or relativistic 
corrections. They are expected to hold at the order $(\frac{1}{m})^0$.
Despite these drawbacks, they roughly agree with some of the more elaborate 
calculations for the first-order crossed diagram \cite{AMGH2,TJON}. 
They however disagree with results of ref. \cite{THEU}. With this respect, 
we notice that the present calculation of each diagram includes some higher 
order effect through the appearance of the effective interaction in its 
expression.  Therefore, the expansion we used for the binding energy 
in terms of crossed diagrams is not equivalent to an expansion in terms 
of the coupling constant, which for a part underlies the results 
of the above work. 

In view of the above remarks and in absence of other calculations, necessarily 
more complicated, we believe that the present results  can give insight on the 
role of higher order terms in the interaction due to crossed diagrams. We 
nevertheless expect some change in the detail, arising from various 
relativistic corrections in particular or other $\frac{1}{m}$ corrections. 
To get information on this, the summation of a subset of crossed diagrams 
should be performed, what supposes to solve a three-body problem 
in the minimal case.

Taking into account the present results as well as those obtained elsewhere in 
the ladder approximation, it appears that the instantaneous approximation 
employed for the usual derivation of interactions is highly misleading. 
The fact that it could effectively work in some cases hides cancellations that 
are in no way small. Moreover, these cases exclude two important physical 
examples: the nucleon-nucleon interaction where the exchanged single pion both 
carries charge and couples to the nucleon spin, and the quark-quark interaction 
where the exchanged gluon carries spin and color charge. Actually, for practical 
applications, it is essential to include the single-boson exchange contribution 
which is essential to describe the long-range part of the force. At shorter 
distances, where the above corrections should show up, the interaction can be 
most usefully fitted to a few experimental data (cross-sections, binding 
energies). As our  study shows,  describing the interaction by single-boson 
exchange in this range, as done in the nucleon-nucleon interaction case, is 
largely elusive. As a support to this practice however, we notice that 
interaction models that completely lack of a theoretical description  in this 
range, like  Argonne V18 \cite{V18}, do quite well.  

\noindent
{\bf Acknowledgments} \\
We are very grateful to A. Amghar for an important observation concerning the 
role of the boson mass in our calculations. 



\begin{thebibliography}{99} 
\bibitem{NIEU} Taco Nieuwenhuis and J.A. Tjon: Phys. Rev. Lett. 
  {\bf 77}, (1996) 814.
\bibitem{BETH} E.E. Salpeter and H.A. Bethe: Phys. Rev. {\bf 84}, (1951) 1232.
\bibitem{WICK} G.C. Wick: Phys. Rev. {\bf 96}, (1954) 1124.
\bibitem{CUTK} R.E. Cutkosky: Phys. Rev. {\bf 96}, (1954) 1135.
\bibitem{SILV} B. Silvestre-Brac et al.: Phys. Rev. {\bf D29}, (1984) 2275.
\bibitem{BILA} A. Bilal and P. Schuck: Phys. Rev. {\bf D31}, (1985) 2045.
\bibitem{PHIL} D.R. Phillips and S.J. Wallace: Phys. Rev. {\bf C54} (1996) 507. 
\bibitem{AMGH} A. Amghar, B. Desplanques, 
               Few-Body Systems {\bf 28} (2000)  65.
\bibitem{AMGH2} A. Amghar, B. Desplanques, and L. Theu{\ss}l,
               Nucl. Phys. {\bf A694} (2001) 439.
\bibitem{TODO} I.T. Todorov: Phys. Rev. {\bf D3}, (1971) 2351.
\bibitem{BREZ} E. Brezin, C. Itzykson  and J. Zinn-Justin: 
  Phys. Rev. {\bf D1}, (1970) 2349.
\bibitem{FRIA} J.L. Friar: Phys. Rev. {\bf C22}, (1980) 796.
\bibitem{GROS} F. Gross: Phys.Rev. {\bf C26}, (1982) 2203.
\bibitem{NEGH} A.R. Neghabian and W. Gl\"ockle: 
  Can. J. Phys. {\bf 61}, (1983) 85.
\bibitem{ITZY} C. Itzykson and J.B. Zuber (eds.): {\em Quantum Field Theory}. 
  McGraw-Hill  International  Editions (1985).
\bibitem{LEPA} G.P. Lepage, nucl-th/9706029.
\bibitem{PARI} M. Lacombe et al., Phys. Rev. {\bf C21} (1980) 861.
\bibitem{BONN} R. Machleidt, K. Holinde and Ch. Elster: Phys. Rep. {\bf 149}, 
  (1987) 1.
\bibitem{THEU} L. Theu{\ss}l and B. Desplanques: 
Few-Body Systems {\bf 30} (2001) 5.
\bibitem{TJON}  J. Tjon, private communication; 
           I.R. Afnan  and  D.R. Phillips, private communication.   
\bibitem{FST}  N. Fukuda K. Sawada M. and Taketani:
  Prog. Theor. Phys. {\bf 12}, (1954) 156. 
\bibitem{OKUB} S. Okubo: Prog. Theor. Phys. {\bf 12}, (1954) 603; \\M. Sugawara 
  and S. Okubo: Phys. Rev. {\bf 117}, (1960) 605.
\bibitem{V18} R.B. Wiringa, V.G.J. Stoks and R. Schiavella: 
   Phys. Rev. {\bf C51} (1995) 38.
\end{thebibliography}
\end{document}